\documentclass{article}
\usepackage[preprint]{spconf}
\copyrightnotice{\begin{minipage}{\textwidth} \copyright\ 2021 IEEE. Personal use of this material is permitted. Permission from IEEE must be obtained for all other uses, in any current or future media, including reprinting/republishing this material for advertising or promotional purposes, creating new collective works, for resale or redistribution to servers or lists, or reuse of any copyrighted component of this work in other works.\end{minipage}}

\usepackage{amsmath,graphicx}

\usepackage{cite}
\usepackage{amsmath,amssymb,amsfonts}
\usepackage{algorithmic}
\usepackage{graphicx}
\usepackage{textcomp}

\usepackage{amsmath,graphicx}
\usepackage{psfrag}
\usepackage{siunitx}
\usepackage{bm,bbm}

\usepackage{tikz,pgfplots}
\usepackage{tikzscale}

\usepackage{rotating}
\usepackage{wrapfig}
\usepackage[linesnumbered,lined,commentsnumbered, ruled]{algorithm2e}

\usepackage[acronym]{glossaries}

\newacronym{DNN}{DNN}{deep neural network}
\newacronym{CDR}{CDR}{coherent-to-diffuse power ratio}
\newacronym{DoA}{DoA}{direction-of-arrival}
\newacronym{TDoA}{TDoA}{time difference of arrival}
\newacronym{MSE}{MSE}{mean-squared error}
\newacronym{MLP}{MLP}{multilayer perceptron}
\newacronym{CRNN}{CRNN}{convolutional recurrent neural network}
\newacronym[plural=GPs,firstplural=Gaussian processes (GPs)]{GP}{GP}{Gaussian process}
\newacronym{GRU}{GRU}{gated recurrent unit}
\newacronym{RIR}{RIR}{room impulse response}
\newacronym{AWGN}{AWGN}{additive white Gaussian noise}
\newacronym{SNR}{SNR}{signal-to-noise ratio}
\newacronym{STFT}{STFT}{short-time Fourier transform}
\newacronym[plural=PSDs,firstplural=power spectral densities~(PSDs)]{PSD}{PSD}{power sprectral density}
\newacronym{AE}{AE}{absolute error}
\newacronym{MAE}{MAE}{mean-absolute error}
\newacronym{PE}{PE}{position error}
\newacronym{MPE}{MPE}{mean position error}
\newacronym{WASN}{WASN}{wireless acoustic sensor network}
\newacronym{OoR}{OoR}{out-of-range}
\newacronym{ASR}{ASR}{automatic speech recognition}
\newacronym{TDNN}{TDNN}{time delay neural network}
\newacronym{CNN}{CNN}{convolutional neural network}
\newacronym{WLS}{WLS}{weighted least squares}
\newacronym{LS}{LS}{least squares}
\newacronym{RANSAC}{RANSAC}{random sample consensus}

\newacronym{GARDE}{GARDE}{\textbf{G}eometry c\textbf{A}libration f\textbf{R}om \textbf{D}istance \textbf{E}stimates}
\newacronym{MDS}{MDS}{Multi Dimensional Scaling}

\newacronym{CWLS}{CWLS}{constrained weighted least squares}
\newacronym{CRLB}{CRLB}{Cramer-Rao lower bound}
\newacronym{RMSE}{RMSE}{root mean square error}

\newacronym{CDF}{CDF}{Cumulative distribution function}
\newacronym{CDF2}{CDF}{Cumulative distribution function}

\newacronym{ABEL}{ABEL}{averaged burst error length}
\newacronym{ACD}{ACD}{average coherence drift}
\newacronym{ADC}{ADC}{analog-digital converter}
\newacronym{APT}{APT}{averaged processing time}
\newacronym{ASN}{ASN}{acoustic sensor network}
\newacronym{ASNs}{ASNs}{acoustic sensor networks}
\newacronym{ASRC}{ASRC}{arbitrary sampling rate conversion}
\newacronym{ATD}{ATD}{time drift}
\newacronym{ATS}{ATS}{accumulating time shift}
\newacronym{BI}{BI}{band-limited interpolation}
\newacronym{BSS}{BSS}{blind source separation}
\newacronym{CCF}{CCF}{cross-correlation function}
\newacronym{CCF-2}{CCF-2}{secondary \gls{CCF}}
\newacronym{CD}{CD}{coherence drift}
\newacronym{CFM}{CFM}{coherence function maximization}
\newacronym{CM}{CM}{correlation maximization}
\newacronym{CSD}{CSD}{cross-spectral density}
\newacronym{CSD-2}{CSD-2}{secondary \gls{CSD}}
\newacronym{CTC}{CTC}{continuous-time conversion}
\newacronym{DCML}{DCML}{\gls{ML} method for dynamic conditions}
\newacronym{DB}{DB}{Data base}
\newacronym{DD}{DD}{digital-to-digital}
\newacronym{DDC}{DDC}{digital-to-digital converter}
\newacronym{DSP}{DSP}{digital signal processing}
\newacronym{DTC}{DTC}{discrete-time conversion}
\newacronym{DXCP}{DXCP}{double-cross-correlation processor}
\newacronym{DXCPPhaT}{DXCP-PhaT}{\gls{DXCP} with phase transform}
\newacronym{ECS}{ECS}{energy correlation score}
\newacronym{FCF}{FCF}{filter correlation function}
\newacronym{FF}{FF}{free-field}
\newacronym{FFT}{FFT}{fast Fourier transform}
\newacronym{FFT-DXCP}{FFT-DXCP}{\gls{FFT} domain \gls{DXCP}}
\newacronym{FO}{FO}{frame-oriented}
\newacronym{GCC}{GCC}{generalized cross-correlation}
\newacronym{GCPSD}{GCPSD}{generalized cross power sprectral density}
\newacronym{GCCF}{GCCF}{generalized \gls{CCF}}
\newacronym{GCCPhaT}{GCC-PhaT}{generalized cross-correlation with phase transform}
\newacronym{GCSD}{GCSD}{generalized cross-spectral density}
\newacronym{GE}{GE}{Gilbert-Elliott}
\newacronym{IBI}{IBI}{iterative \gls{BI}}
\newacronym{ICA}{ICA}{independent component analysis}
\newacronym{IML}{IML}{iterative \gls{ML}}
\newacronym{IR}{IR}{impulse response}
\newacronym{IIR}{IIR}{infinite impulse response}
\newacronym{IFFT}{IFFT}{inverse \gls{FFT}}
\newacronym{ISTFT}{ISTFT}{inverse \gls{STFT}}
\newacronym{LCD}{LCD}{least-squares coherence drift}
\newacronym{LPD}{LPD}{linear-phase drift}
\newacronym{LTI}{LTI}{linear time-invariant}
\newacronym{LTV}{LTV}{linear time-variant}
\newacronym{LTVIR}{LTV-IR}{linear time-variant impulse response}
\newacronym{ML}{ML}{maximum likelihood}
\newacronym{MS}{MS}{multi-stage}
\newacronym{MSC}{MSC}{magnitude-squared coherence}
\newacronym{OML}{OML}{optimized \gls{ML}}
\newacronym{OR}{OR}{outlier removal}
\newacronym{PHAT}{PHAT}{phase transform}
\newacronym{PL}{PL}{packet loss}
\newacronym{PolyFar}{PolyFar}{polyphase-Farrow}
\newacronym{ppb}{ppb}{part per billion}
\newacronym{ppm}{ppm}{parts \ per \ million}

\newacronym{RBI}{RBI}{recursive band-limited interpolation}
\newacronym{RF}{RF}{realtime factor}
\newacronym{RTP}{RTP}{Real-time Transport Protocol}
\newacronym{RTCP}{RTCP}{Real-time Transport Control Protocol}

\newacronym{SCOT}{SCOT}{smoothed coherence transform}
\newacronym{SINR}{SINR}{signal-to-interpolation-noise ratio}
\newacronym{SO}{SO}{sample-oriented}
\newacronym{SPIB}{SPIB}{signal processing information base}
\newacronym{SPL}{SPL}{signal packet loss}
\newacronym{SRC}{SRC}{sampling rate conversion}
\newacronym{SRO}{SRO}{sampling rate offset}
\newacronym{STO}{STO}{sampling time offset}
\newacronym{TCP}{TCP}{Transmission Control Protocol}
\newacronym{TD}{TD}{time domain}
\newacronym{TDDXCP}{TD-DXCP}{time domain \gls{DXCP}}
\newacronym{TDD}{TDD}{time-delay difference}
\newacronym{TDOA}{TDOA}{time-difference of arrival}
\newacronym{UDP}{UDP}{User Datagram Protocol}
\newacronym{VAD}{VAD}{activity detection}
\newacronym{SAD}{SAD}{sound activity detection}
\newacronym{WACD}{WACD}{weighted average coherence drift}
\newacronym{DWACD}{DWACD}{dynamic weighted average coherence drift}
\newacronym{WG}{WG}{weighting}
\newacronym{WLCD}{WLCD}{weighted \gls{LCD}}
\newacronym{WASNs}{WASNs}{wireless acoustic sensor networks}
\newacronym{WCP}{WCP}{wideband correlation processor}
\newacronym{XC}{XC}{cross-correlation}

\newacronym{TXCO}{TXCO}{temperature compensated crystal oscillator}

\newacronym{OU}{OU}{Ornstein-Uhlenbeck}

\newacronym[firstplural=time differences of flight (TDOFs)]{TDOF}{TDOF}{time difference of flight}

\usepackage[colorlinks]{hyperref}
\usepackage{xcolor}
\usepackage{import}
\usepackage{booktabs}
\usepackage{mathtools}
\usepackage{microtype}
\usepackage[autostyle=true]{csquotes}
\usepackage{cleveref}
\usepackage{multirow}
\usepackage{pgfplots}
\usepackage[normalem]{ulem}
\usepackage{amssymb}
\usepackage{pifont}
\newcommand{\cmark}{\ding{51}}%
\pgfplotsset{compat=newest}
\usepgfplotslibrary{groupplots}
\usepgfplotslibrary{dateplot}
\usepackage[symbol]{footmisc}

\newcommand{\taugeo}[2]{\tau^{(#1)}_{#2}}
\newcommand{\tausto}[1]{\tau^\text{STO}_{#1}}
\newcommand{\eststo}[1]{\hat{\tau}^\text{STO}_{#1}}

\newcommand{\rir}[3]{h_{#1}^{(#2)}(#3)}

\newcommand{\srcsig}[2]{x^{(#1)}(#2)}
\newcommand{\noise}[2]{v_{#1}(#2)}
\newcommand{\micsig}[2]{y_{#1}#2}

\newcommand{\sro}[2]{\varepsilon_{#1}[#2]}
\newcommand{\avgsro}[2]{\bar{\varepsilon}_{#1}\! \left[#2\right]}
\newcommand{\estsro}[2]{\widehat{\varepsilon}_{#1}\! \left[#2\right]}
\newcommand{\delay}[1]{T_{#1}}

\newcommand{\freq}[2]{f_#1[#2]}
\newcommand{\totaldelay}[2]{\tau_#1[#2]}
\newcommand{\avgtotaldelay}[2]{\bar{\tau}_#1[#2]}
\newcommand{\dist}[2]{d_{#1}^{(#2)}}
\newcommand{\estdist}[3]{\hat{d}_{#1}^{(#2)}[#3]}
\newcommand{\cp}[1]{P_\Gamma(#1)}

\newcommand{\wacd}[1]{P_\text{WACD}(#1)}
\newcommand{\wacdtime}[1]{p_\text{WACD}(#1)}
\newcommand{\tauwacd}[2]{\bar{\tau}^{\text{CD}}_{#1}[#2]}
\newcommand{\taucomp}[2]{\tau^{\text{c}}_{#1}[#2]}
\newcommand{\tmpdist}{\ell_d}

\newcommand{\esttaugescomp}[2]{\hat{\tau}_{#1}[#2]}

\newcommand{\epa}[1]{\hat{\varepsilon}_{12}[\tilde{\ell'}]}

\usepackage{tikzscale}

\title{On Synchronization of wireless acoustic sensor networks in the presence of time-varying sampling rate offsets and speaker changes}
\name{Tobias Gburrek, Joerg Schmalenstroeer, Reinhold Haeb-Umbach}
\address{Department of Communications Engineering, Paderborn University, Germany \\ \{gburrek, schmalen, haeb\}@nt.uni-paderborn.de}

\usepackage[skip=5pt]{caption}
\begin{document}
	\ninept
	\setlength{\textfloatsep}{3pt}
	\setlength{\intextsep}{3pt}
	\setlength{\intextsep}{2pt}
	\setlength{\abovedisplayskip}{2pt}
	\setlength{\abovedisplayshortskip}{2pt}
	\setlength{\belowdisplayskip}{2pt}
	\setlength{\belowdisplayshortskip}{2pt}

	\maketitle
	\begin{abstract}
		A wireless acoustic sensor network records audio signals with sampling time and sampling rate offsets between the audio streams, if the \glspl{ADC} of the network devices are not synchronized. 
		Here, we introduce a new sampling rate offset model to simulate time-varying sampling frequencies caused, for example, by temperature changes of \gls{ADC} crystal oscillators, and propose an estimation algorithm to handle this dynamic aspect in combination with changing acoustic source positions. Furthermore, we show how \acrlong{DNN} based estimates of the distances between microphones and human speakers can be used to determine the sampling time offsets. This enables a synchronization of the audio streams to reflect the physical time differences of flight.
	\end{abstract}
	\begin{keywords}
		Synchronization, time-varying sampling rate offset, sampling time offset
	\end{keywords}

	\glsresetall

\section{Introduction}
\label{sec:intro}

\Gls{SRO} and \gls{STO} estimation have received a lot of interest in recent years \cite{Markovich-Golan2012, Miyabe2013, Bahari2017, SchHaeb2017, Araki19, Chinaev2021}. In particular, in the context of \glspl{WASN} these topics are important, because for many signal processing tasks, such as acoustic beamforming \cite{Gannot2016Consolidated}, the audio signals must be synchronized.

\Glspl{SRO} are caused by the differences in the actual frequencies of the oscillators driving the \glspl{ADC} of the distributed devices despite their equal nominal frequencies \cite{SchJebHaeb2014}. The author of \cite{DRIFT2021}  showed that typical \gls{SRO} values range between \SI{-500}{\gls{ppm}} and \SI{+400}{ppm} from the nominal frequency for handheld devices, such as smartphones. \Glspl{STO} arise due to the fact that the devices start recording at different moments in time. 

The vast majority of works on \gls{SRO} and \gls{STO} estimation from the recorded audio signals are conducted under the assumptions of constant \glspl{SRO} and fixed positions of sources and sensors, which, however, are unrealistic for many practical scenarios.
The frequency of a crystal oscillator is not only determined by the crystal's shape and the technique utilized to cut it, but is also influenced by a variety of environmental factors, such as aging, temperature, humidity and supply voltage \cite{Walls92}. While constant or slowly changing factors, like aging, are sufficiently accurately modeled by assuming a constant \gls{SRO}, temperature and supply voltage can change more rapidly and challenge the assumption of constant \glspl{SRO} \cite{SchJebHaeb2014}. In \cite{QSL2021} the authors noted that when the device is switched on or when it transits from sleeping to processing mode, the change of the device's temperature incurs a sampling rate change of several ppm  within a relatively short period of time (a few minutes). But even after warm-up the \gls{SRO} can be time-varying, e.g., due to fluctuations of the supply voltage, which are caused by the changing workload of the microprocessor \cite{SchJebHaeb2014}. This all stands in conflict with the assumption of constant \glspl{SRO} for relatively long time intervals of, e.g., several minutes, made by several \gls{SRO} estimation algorithms.


Likewise, the assumption of a fixed position of the acoustic source, that is made by most synchronization algorithms that estimate the \gls{SRO} from the acoustic signals received by the devices,  is not given in many application scenarios of a \gls{WASN}. Consider, for  example, a meeting where different participants, each speaking at a different position relative to the sensor network. 
Handling moving speakers would be even more complicated.
Estimating an \gls{SRO} from signals emitted by a moving source faces the challenging task of separating the \gls{SRO}-induced delays from delays caused by source position changes. A reasonable simplification is to focus on time periods where the positions remain fixed and to skip transient phases, as proposed in \cite{Araki19}. 

In this contribution we attempt to overcome some of the above mentioned assumptions. First, we will model the time-varying \glspl{SRO} with an Ornstein-Uhlenbeck process \cite{Uhlenbeck} and then devise an algorithm for \gls{SRO} estimation. We discuss the necessary modifications to a previously proposed coherence drift based online algorithm~\cite{Chinaev2021} to deal with source position changes and time-varying \glspl{SRO}. Second, we consider \gls{STO} estimation by separating time delay estimates into the contribution of the recording start times  and the contribution from the \glspl{TDOF} from the speakers' positions to the sensors. The latter requires an estimate of the distance of the speaker to the microphones, which is obtained from a \gls{DNN} based distance estimator \cite{Gburrek21b}. By discriminating these different contributions we arrive at an \gls{STO} estimate that regards the physical \glspl{TDOF}. This is important, e.g., if the location of the sensor nodes is to be estimated from \gls{TDOA} estimates.

The paper is organized as follows: In Sec.~\ref{sec:problem} we discuss the impact of time-varying \glspl{SRO} on signals and show in Sec.~\ref{sec:wacd} our modifications to handle dynamic scenarios with source position changes and time-varying \glspl{SRO}  during \gls{SRO} estimation.
Subsequently, \gls{STO} estimation is presented in Sec.~\ref{sec:sto}. Our approach for modeling time-varying \glspl{SRO} is explained in Sec.~\ref{sec:sro_model}, before we discuss the experiments in Sec.~\ref{sec:exp} and end with the conclusions drawn in Sec.~\ref{sec:conclusions}.


\section{Problem Statement}
\label{sec:problem}
We consider a \gls{WASN} consisting of two sensor nodes and an unknown \gls{SRO} and \gls{STO} between them. Each sensor node is equipped with at least two microphones whose signals are synchronously sampled.
Furthermore, we assume that both sensor nodes record the signals emitted by sources at $M$ stationary positions whereby at most one source is active at any given time.
Additionally, we consider that there might be short periods of time without source activity when the source position changes.
Such a scenario could be typical of a meeting with multiple participants sitting around a conference table with multiple microphone arrays in the middle.

To simplify the presentation,  only the first microphone channel of each sensor node will be reflected in the notation.
The continuous-time microphone signal of the $i$-th sensor node, with $i \in \{1, 2\}$, is given by
\begin{align}
	\micsig{i}{(t)} = \sum_{m=1}^{M} \rir{i}{m}{t} \ast \srcsig{m}{t} + \noise{i}{t},
\end{align}
with $\srcsig{m}{t}$ being the source signal emitted at the $m$-th position and $\noise{i}{t}$ representing white Gaussian sensor noise.
$\rir{i}{m}{t}$ denotes the \gls{RIR} modeling the sound propagation from the $m$-th source position to the position of the $i$-th microphone.

Typically, the sampling processes of the microphone signals will start at different points in time resulting in the \gls{STO} $T_i$.
In addition to that, the sampling frequency of the $i$-th microphone $\freq{i}{n} {= }\left(1 + \sro{i}{n}\right) \cdot f_s$ will slightly deviate from the nominal sampling frequency $f_s$.
Here, $n$ denotes the discrete-time sample index and $\sro{i}{n}$ the time-varying \gls{SRO} of the $i$-th microphone with $\left|\sro{i}{n}\right| {\ll} 1$.
Sampling the $i$-th microphone signal results in the following discrete-time signal (with $1 / (1+x) \approx 1 - x $ if $|x| \ll 1$):
\renewcommand{\thefootnote}{(\fnsymbol{footnote})}
\begin{align}
\micsig{i}{[n]} &= \micsig{i}{\left(\delay{i} + \sum_{\tilde{n}=0}^{n-1}\frac{1}{f_s\cdot(1+\sro{i}{\tilde{n}})}\right)}\nonumber \\ 
&\approx \micsig{i}{\left(\frac{n}{f_s} - \frac{1}{f_s} \cdot \left(\smash[b]{\underbrace{-\delay{i} \cdot f_s + \sum_{\tilde{n}=0}^{n-1} \sro{i}{\tilde{n}}}_{\coloneqq \totaldelay{i}{n}}}\right)\right)}.
\vphantom{\underbrace{+ \delay{i} - \sum_{\tilde{n}=0}^{n-1} \frac{\sro{i}{\tilde{n}}}{f_s}}_{\coloneqq - \totaldelay{i}{n} / f_s}}
\end{align}
Compared to the synchronous signal $y_i^{(\text{sync})}[n]$ being sampled with the nominal sampling frequency $f_s$ ($\sro{i}{n}{=}\SI{0}{\gls{ppm}}$ and $T_i{=}\SI{0}{\second}$) $\micsig{i}{[n]}$ shows a shift of $\totaldelay{i}{n}$ \si{samples}.
\setcounter{footnote}{1}

Usually, the microphone signals are processed in a frame-oriented fashion, e.g., in the \gls{STFT} domain.
If the shift $\totaldelay{i}{n}$ is much smaller than the \gls{STFT} frame size $N$ the relation between the asynchronously sampled signal $y_i[n]$ and the synchronously sampled signal $y_i^{(\text{sync})}[n]$ can be modeled in the \gls{STFT}-domain as follows~\cite{wang16,Bahari2017,SchHaeb2017}:
\begin{align} \label{EQ:SRO_Phase_Mult}
	Y_i(l B, k) \approx Y^{(\text{sync})}_\text{i}(l B, k)  \cdot \exp{\left(-j\frac{2 \pi k}{N} \avgtotaldelay{i}{l}\right)},
\end{align}
with frame index $l$, frequency bin index $k$ and frame shift $B$.
In addition, the \gls{SRO} is assumed to change slowly and is therefore approximated by $\avgsro{i}{l}$ during a \gls{STFT} frame.
Hence, the average shift of the $l$-th frame is given by
\begin{align} \label{EQ:SRO_ACC}
\avgtotaldelay{i}{l} =   -\delay{i}\cdot f_s +\left( \frac{N}{2} \avgsro{i}{0} + \sum_{\tilde{l}=1}^{l} \avgsro{i}{\tilde{l}} \cdot B\right).
\end{align}

Without loss of generality, the first microphone is selected as reference and the $m$-th position is regarded in the following.
The task at hand can now be described as follows: Estimate the \gls{SRO} $\avgsro{12}{l} {\approx} \avgsro{2}{l}-\avgsro{1}{l}$ and the \gls{STO} $\tausto{12} {=} (\delay{2}-\delay{1}) \cdot f_s$. Furthermore, we wish to distinguish  between the contribution of the \gls{STO} $\tausto{12}$  and the \gls{TDOF} $\taugeo{m}{12} {=} (\dist{2}{m} - \dist{1}{m})/c \cdot f_s$ to the overall signal shift with  $c$ denoting the speed of sound and $\dist{i}{m}$ the distance between the $i$-th microphone and the source at the $m$-th position.


\section{Dynamic Weighted Average Coherence Drift}
\label{sec:wacd}
In this section, we briefly recapitulate the concept of our previously proposed online \gls{WACD} method from  \cite{Chinaev2021} and describe how it is adapted to the considered dynamic scenario, resulting in the \gls{DWACD} method.




Online \gls{WACD} estimates the coherence $\smash{\Gamma^{\taucomp{12}{\ell}}_{12}(\ell B_s, k)}$ segment-wisely, where $\ell$ represents the segment index and $B_s$ the segment shift, and where a segment comprises several frames. Thereby, $\taucomp{12}{\ell}$ corresponds to an adjustable shift of the segment taken from $y_2[n]$ that coarsely compensates for the \gls{SRO}-induced signal shift \cite{Chinaev2021}.
The coherence is calculated from microphone signal segments of length $N_W$, where the $\ell$-th segment taken from the first microphone starts at $\micsig{1}{[\ell B_s]}$ and the $\ell$-th segment taken from the second microphone at $\micsig{2}{[\ell B_s + \taucomp{12}{\ell}]}$.
It is to be mentioned that the \gls{SRO} is also estimated every $B_s$ \si{samples}.

As shown in~\cite[Eq. (23)]{SchHaeb2017} the coherence is given by:
\begin{align}
	\Gamma^{\taucomp{12}{\ell}}_{12}(\ell B_s, k) {\approx } H_{12}^{(m)}(k) {\cdot} W_\text{SNR}(\ell, k) {\cdot} \exp{\left(j\frac{2 \pi k}{N}\tauwacd{12}{\ell}\right)}, 
\end{align}
with $H_{12}^{(m)}(k) {=} H_{1}^{(m)}(k) {\cdot} (H_{2}^{(m)}(k))^*$ and $W_\text{SNR}(\ell, k)$ being a weight depending on the \gls{SNR} of the time-frequency bins.
$\tauwacd{12}{\ell}$ corresponds to the average shift of the $\ell$-th segment:
\begin{align}
\tauwacd{12}{\ell} {=} {-} \tausto{12} {-} \taucomp{12}{\ell} {+}  \left(\frac{N_W}{2} \avgsro{12}{0} {+} \sum_{\tilde{\ell}=1}^{\ell} \avgsro{12}{\tilde{\ell}} {\cdot} B_s\right).
\end{align}

The \gls{SRO} is estimated based on the complex conjugated product of consecutive coherence functions with a temporal distance of $\tmpdist B_s$ samples:
\begin{align}
\cp{\ell, k} = \Gamma^{\taucomp{12}{\ell}}_{12}(\ell B_s, k) \cdot \left(\Gamma^{\taucomp{12}{\ell}}_{12}((\ell-\tmpdist) B_s, k)\right)^*.
\end{align}
Taking into account that the \gls{SRO} changes only slowly, the \gls{SRO} can be approximated by $\avgsro{12}{\ell}$ in the interval $[(\ell - \tmpdist) B_s,  \ell B_s]$ resulting in (see~\cite[Eq. (14)]{SchHaeb2017})
\begin{align}
\cp{\ell,k} \approx W(\ell, k) \cdot \exp{\left(j \frac{2\pi k}{N}  \tmpdist B_s \avgsro{12}{\ell} \right)},
\end{align}
whereby $W(\ell, k){=}\left|H_{12}^{(m)}(k)\right|^2 {\cdot} W_\text{SNR}(\ell, k){\cdot}W_\text{SNR}^*(\ell-\ell_d, k)$ corresponds to  an~\gls{SNR}-related weight.

The \gls{DWACD} method which is presented in the following utilizes a temporally averaged version of the complex conjugated product of consecutive coherence functions $\cp{\ell,k}$ to estimate the \gls{SRO}:
\begin{align}
	\label{eq:smoothed_P_wacd}
\wacd{\ell,k} = \alpha \cdot \wacd{\ell-1,k} + (1-\alpha) \cdot \cp{\ell,k},
\end{align}
with $\wacd{-1,k} {=}0$ as initial value and $\alpha$ being a smoothing factor close to $1$.
In contrast to the online \gls{WACD} method, the \gls{DWACD} method uses a temporal weighting by the autoregressive smoothing to be able to adapt to fast \gls{SRO} changes by giving more recent estimates $\cp{\ell,k}$ a larger weight.
Using an energy-based \gls{SAD}, the weighted average coherence product is only updated if a source is active in all signal segments needed to calculate $\smash{\Gamma^{\taucomp{12}{\ell}}_{12}(\ell B_s, k)}$ and $ \smash{\Gamma^{\taucomp{12}{\ell}}_{12}((\ell - \tmpdist)B_s, k)}$.

A crucial assumption for coherence drift based \gls{SRO} estimation is that the weight $W(\ell, k)$ does not contribute to the phase of $\cp{\ell,k}$~\cite{SchHaeb2017}, which in consequence means that the source position has to be constant during the interval used to compute $\cp{\ell,k}$. Thus, in a scenario with changing source positions, the \gls{DWACD} method has to use segments with smaller lengths $N_W$ and smaller time interval $\tmpdist$ between successive coherence functions to avoid or to at least reduce the probability of a speaker change in the estimation interval.

Moreover, the \gls{SRO} would have to be zero or at least close to zero for the phase of the weight $W(\ell, k)$ to be close to zero~\cite{SchHaeb2017}. The online~\gls{WACD} method does not specifically address this issue while the offline \gls{WACD} method as proposed in~\cite{SchHaeb2017} utilizes a multi-stage approach with resampling to solve it.
The \gls{DWACD} method takes the \gls{SRO} estimate of the previous segment $\estsro{12}{\ell-1}$ to resample the segment of $y_2[n]$ which is currently used for coherence estimation.
This resampling is realized by a multiplication of the $\kappa$-th \gls{STFT}-frame of the Welch method used for coherence estimation with the phase term $\exp{\left(j \frac{2\pi k}{N}  \kappa B \estsro{12}{\ell-1} \right)}$.

The online \gls{WACD} method estimates the \gls{SRO} from the phase of $\wacd{\ell,k}$, which suffers from the $2\pi$-periodicity of the phase and outlier time-frequency bins, i.e., bins with exceptionally large phase errors.
Such outlier time-frequency bins occur more frequently in the multi-position scenario where shorter averaging periods, shorter segments and  shorter temporal distances $\tmpdist$ have to be used to calculate $\wacd{\ell,k}$.
Interpreting $\wacd{\ell,k}$ as a \gls{GCPSD}~\cite{knapp76} with \gls{SNR}-based weights, \gls{DWACD} estimates the \gls{SRO} using the time lag $\lambda_\text{max}$ that maximizes the \gls{GCC} function $\wacdtime{\ell, \lambda}$:
\begin{align}
	\estsro{12}{\ell} = -\frac{1}{\tmpdist B_s} \cdot \lambda_\text{max} =- \frac{1}{\tmpdist B_s} \cdot \underset{\lambda}{\text{argmax}} \ |\wacdtime{\ell, \lambda}|,
\end{align}
with $\wacdtime{\ell, \lambda} = \text{IFFT}\{\wacd{\ell,k}\}$ as the $N$-point IFFT.

For an accurate \gls{SRO} estimation a golden section search in the interval $[\lambda_\text{max}{-}0.5,\lambda_\text{max}{+}0.5]$ is used to find the non-integer time lag $\lambda \ \in \ \mathbb{R}$ that maximizes $|\wacdtime{\ell, \lambda}|$.
Note that a settling time is introduced, i.e., the \gls{SRO} is only estimated for $\ell \geq 40$, to guarantee that $\wacd{\ell,k}$ from Eq.~\eqref{eq:smoothed_P_wacd} is settled.
As discussed in~\cite{Chinaev2021}, a coarse synchronization of audio streams is necessary for accurate \gls{SRO} estimation. Consequently, the integer offset between the microphone signals is determined during the first \SI{20}{\second} of source activity using a cross-correlation, and subsequently compensated before applying \gls{DWACD}.
We propose to use the following parameters: \gls{STFT} using a Blackman window; $B{=}2^9$; $N{=}2^{12}$; $B_s{=}2^{11}$; $N_W{=}2^{13}$; $\tmpdist=4$, $\alpha=0.95$.


\section{Sampling Time Offset Estimation}
\label{sec:sto}
In this section a method for \gls{STO} estimation is proposed that utilizes the \gls{SRO} compensated signal of $y_2[n]$. The remaining unknown shift between the microphone signals for the $m$-th source position is given by (see~\eqref{EQ:SRO_ACC} with an additional shift corresponding to the \gls{TDOF})
\begin{align}
	\tau_{12} = \taugeo{m}{12}   - \tausto{12} = \frac{\dist{2}{m} - \dist{1}{m}}{c} \cdot f_s  - \tausto{12}.
\end{align}


The shift $\tau_{12}$ and the distances $\dist{1}{m}$ and $\dist{2}{m}$ needed for \gls{STO} estimation are segment-wisely estimated with a segment length of $2^{14}$~\si{samples} and a segment shift of $2^{11}$~\si{samples}.
The \gls{GCCPhaT} algorithm is used to gather estimates $\hat{\tau}_{12}[\ell]$ for the remaining signal shift.
Furthermore, the \gls{DNN}-based distance estimator~\cite{Gburrek21b} with \gls{CDR} and \gls{STFT} as input features is used to estimate the distances $\estdist{1}{m}{\ell}$ and $\estdist{2}{m}{\ell}$ of the source to the two sensor nodes.


Based on these estimates, a \gls{LS} problem minimizing the error $\sum_{\ell=0}^{L-1} \!(\tausto{12} {-} (\estdist{2}{m}{\ell} {-} \estdist{\ell}{m}{l})/c \cdot f_s {+} \esttaugescomp{12}{\ell})^2$ is solved for \gls{STO} estimation.
Hereby, $L$ is the number of considered segments.
Segments without source activity are excluded from the \gls{LS} problem using an energy-based \gls{SAD}.
The \gls{LS} problem leads to the following \gls{STO} estimate:
\begin{align}
\eststo{12} = \frac{1}{L}\sum_{\ell=0}^{L} \!\left(\frac{\estdist{2}{m}{\ell} {-} \estdist{1}{m}{\ell}}{c} \cdot f_s {-} \esttaugescomp{12}{\ell}\right).
\end{align}
Due to the fact that the shift estimates $\esttaugescomp{12}{\ell}$ as well as the distance estimates $\estdist{1}{m}{\ell}$ and $\estdist{2}{m}{\ell}$ may exhibit large errors for specific constellations of source position and sensor node positions we embed the LS solver in a \gls{RANSAC}~\cite{fischler81} method to remove outliers from the estimation.

\section{Dynamic SRO model}
\label{sec:sro_model}
\begin{figure}[tb]
	\centering
	\Huge
	\pgfplotsset{/pgfplots/group/.cd,
		horizontal sep=4.25cm
	}
	\resizebox{\linewidth}{!}{
		\input{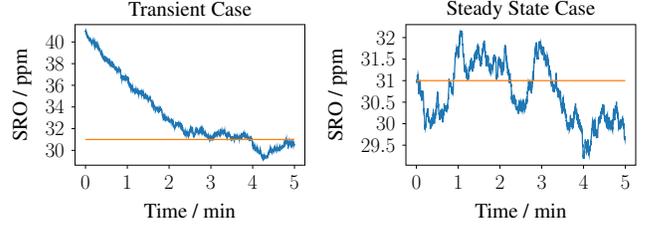}
	}
	\caption{Example \gls{SRO} trajectories for a transient case ($\mu_\infty{=}31$ppm; $\epsilon[0]{=}\mu_\infty{+}10\text{ppm}$) and steady state case ($\mu_\infty{=}31$ppm; $\epsilon[0]{=}\mu_\infty$).} \label{Fig:cd}
\end{figure}

In the experiments we model fluctuations of the \gls{SRO} by an Ornstein-Uhlenbeck process \cite{Uhlenbeck}. This process is flexible enough to model both initial transients in \gls{SRO}, e.g., caused by warm-up after switching on the device, and steady-state fluctuations, e.g., due to changing workload of the processor. The Ornstein-Uhlenbeck process is implemented as an auto-regressive process using the discrete-time Euler-Maruyama approximation: 
\begin{align} \label{EQ:EULERMARU}
\varepsilon[\ell] = \varepsilon[\ell-1] + \theta \cdot \left(\mu_\infty - \varepsilon[\ell-1] \right) + x_{\varepsilon}[\ell],
\end{align}
with smoothing factor $\theta {\ll} 1$,  $x_{\varepsilon}[\ell] \sim \mathcal{N}(0; \sigma_\text{OU}^2)$ drawn from a zero-mean Gaussian distribution with variance $\sigma_\text{OU}^2$ and $\mu_\infty$ being the mean \gls{SRO} value reached after all transient effects have died out. In the experiments we set $\sigma_\text{OU} = \SI{0.05}{ppm}$ resulting in a steady state standard deviation of the \gls{SRO} of \SI{1.25}{ppm} and limit the \gls{SRO} range for $\mu_\infty$ to $\pm \SI{100}{ppm}$, which excludes the extreme values reported in \cite{DRIFT2021}.
The start value is chosen to be $\varepsilon[0] = \mu_\infty + \Delta_\text{start}$ with $\Delta_\text{start}$ in the range $\pm \SI{10}{ppm}$.
Fig.~\ref{Fig:cd} shows two exemplary \gls{SRO} trajectories: The left plot displays a transient example simulating a device with a temperature change and the right example showcases a device in a steady state.  

Note, if the circuit driving the \gls{ADC} is using a \gls{TXCO} the temperature dependent transient effect is marginalized ($\varepsilon[0] := \mu_\infty$) while the other influences remain \cite{TCXO14}, e.g., the SiT5156 achieves a temperature stability of $\SI{\pm 0.5}{ppm}$ and has an overall stability of $\SI{\pm 2.5}{ppm}$.



\section{Experiments}
\label{sec:exp}
\renewcommand{\thefootnote}{\arabic{footnote}}
On a simulated data set, we evaluate the presented \gls{STO} estimator and the \gls{DWACD} approach, comparing the latter with two state-of-the-art online \gls{SRO} estimators\footnote[1]{Source code is available at {\scriptsize \url{https://github.com/fgnt/paderwasn}}}.
%
%
The data  is based on the \gls{RIR} data set (reverberation time $T_{60} {=} \SI{300}{\milli\second}$) utilized for geometry calibration in~\cite{gburrek2021geometry}.
Using this data set $100$ \glspl{WASN} with four nodes were simulated whereby the first node was used as reference node for \gls{SRO} and \gls{STO} estimation.
The sensor nodes start recording with an \gls{STO} in the range $\pm \SI{1}{\second}$.
Each recording is \SI{5}{\minute} long and contains a random number of positions $M$ at which up to 4 utterances from the TIMIT data base~\cite{timit} are used as source signals.
Speech pauses with a length between \SI{0.5}{\second} and \SI{2}{\second} are optionally added during the source position changes.
For the generation of  signals with time-varying \glspl{SRO}, the \gls{STFT}-resampling method from~\cite{Schmalen2018efficient} was utilized.
Sensor noise was added  such that the \gls{SNR} has an average value of \SI{30}{\decibel} for a source-node distance of \SI{3.2}{\metre}, which is  the average  distance on the data set.

\begin{table}[tb]
	\centering
	\setlength{\tabcolsep}{4pt}
	\caption{Average and maximum RMSE of the \gls{SRO} estimates and the resulting RMSE of the \gls{SRO}-induced delay estimates $\tau^\varepsilon_{12}$}
	\begin{tabular}{l|c c c|c c c c}
		\toprule	
		& \multirow{4}{*}{\rotatebox{90}{$\bm{\varepsilon \neq} $ \textbf{const.}}} & \multirow{4}{*}{\rotatebox{90}{{\textbf{Multi-Pos.}}}}& \multirow{4}{*}{\rotatebox{90}{\textbf{Silence}}} & \multirow{4}{*}{\textbf{Method}} &\textbf{avg.}  &\textbf{avg.} &\textbf{max.}\\
		& & &  &   & \textbf{RMSE} & \textbf{RMSE} & \textbf{RMSE}\\
		& & &  &   & $\bm{\varepsilon_{12} }$ & $\bm{\tau^\varepsilon_{12}}$ &  $\bm{\tau^\varepsilon_{12}}$ \\
		& & & &   &\textbf{ / \si{\textbf{ppm}}} & \textbf{/ \si{\textbf{samples}}}& \textbf{/ \si{\textbf{samples}}}\\	
		\midrule
		\multirow{4}{*}{\rotatebox{90}{\textbf{Scenario-1}}} &  & & & Online WACD &0.21 &0.14 &0.50 \\
		&  &  &  & DXCP-PhaT &0.15 &0.36 &0.68 \\
		&  &  &  & DXCP-PhaT$_8$ &0.69 &1.73 &3.65 \\
		&  &  &  & DWACD &0.40 &0.15 &0.50\\
		\midrule
		\multirow{4}{*}{\rotatebox{90}{\textbf{Scenario-2}}} &\cmark  & &  &  Online WACD &0.63 &0.73 &2.09 \\
		&\cmark  & &  & DXCP-PhaT &0.66 & 0.97 &2.73 \\
		&\cmark  & &  & DXCP-PhaT$_8$ &0.95 &1.83 &4.66 \\
		&\cmark  & &  & DWACD &0.51  & 0.27 &1.04\\
		\midrule
		\multirow{4}{*}{\rotatebox{90}{\textbf{Scenario-3}}} &\cmark  &\cmark &  \cmark&  Online WACD &2.98 &6.04 &21.00 \\
		&\cmark &\cmark & \cmark & DXCP-PhaT &28.96 &21.84 &161.54 \\
		& \cmark &\cmark & \cmark & DXCP-PhaT$_8$ &1.31 &2.70 &7.76 \\
		& \cmark &\cmark & \cmark & DWACD &0.57 &0.32 &1.20 \\
		\midrule
		\multirow{4}{*}{\rotatebox{90}{\textbf{Scenario-4}}} &\cmark  &\cmark  & &  Online WACD &2.80 &3.25 &10.96\\
		& \cmark &\cmark   &  & DXCP-PhaT &22.42 &16.61 &160.49 \\
		& \cmark &\cmark &  & DXCP-PhaT$_8$ &1.28 &2.81 &6.93\\\texttt{}
		& \cmark &\cmark  & & DWACD &0.64 &0.32 &1.10 \\
		\bottomrule
	\end{tabular}
	\label{tab:comparison_sro}
\end{table} 

\begin{table}[b]
	\centering
	\caption{Dependency of the average RMSE  / \si{ppm} of the \gls{SRO} estimates on the standard deviation $\sigma_\varepsilon$ of $\varepsilon_{12}[\ell]$ for \textit{Scenario-2}} 
	\begin{tabular}{ c c c c }
		\toprule	
		$\bm{\sigma_\varepsilon}$\textbf{ / \si{\textbf{ppm}}} & \textbf{Online WACD}& \textbf{DXCP-PhaT} &\textbf{DWACD} \\
		\midrule
		0 - 1 &0.55 &0.54 &0.49\\
		1 - 2 &0.60 &0.61 &0.51 \\
		2 - 3 &0.63 &0.69 &0.51 \\
		3 - 4 &0.71 &0.80 &0.54 \\
		\bottomrule
	\end{tabular}
	\label{tab:dyn_sro}
\end{table}

In Tab.~\ref{tab:comparison_sro} and Tab.~\ref{tab:dyn_sro} the proposed \gls{DWACD} method is compared to the online \gls{WACD} method we presented in~\cite{Chinaev2021} and the DXCP-PhaT algorithm~\cite{Chinaev21_DXCPPhaT}.
For all algorithms the signals were initially coarsely synchronized as described in Sec.~\ref{sec:wacd} before \gls{SRO} estimation.
The average over the last $160$ complex conjugated coherence products is used for \gls{SRO} estimation in the online \gls{WACD} method.
Due to the fact that the temporal distance between the two signal segments used to calculate the secondary \gls{GCPSD} function is quite long  in the  original DXCP-PhaT ($\approx$ \SI{5}{\second}) and inappropriate for the considered scenario with source position changes, the algorithm is also evaluated with a reduced temporal distance of $8$ \gls{STFT} frames ($\approx$ \SI{1}{\second}) (denoted as DXCP-PhaT$_8$).

Tab.~\ref{tab:comparison_sro} shows the \gls{SRO} error for four scenarios with different degrees of dynamicity.
Further, the average and maximum error of the \gls{SRO}-induced delay $\tau^\varepsilon_{12}$, which is calculated as described in Eq.~\eqref{EQ:SRO_ACC}, are shown.
In \textit{Scenario-1} with constant \gls{SRO} and a single source position all methods are able to deliver precise \gls{SRO} estimates.
Hereby, the online \gls{WACD} method and DXCP-PhaT which were designed for such a setup show the best performance.
In \mbox{\textit{Scenario-2}} a time-varying \gls{SRO} is considered. Here, the performance of all estimators degrades, with the \gls{DWACD} method exhibiting the least degradation.

Taking into account source positions changes (\textit{Scenario-3} and \textit{Scenario-4}) the error grows a lot for the online \gls{WACD} method and DXCP-PhaT while the error for the  \gls{DWACD} method stays nearly on the same level as for a fixed source position.
It is evident, that  position changes are more detrimental to the estimation performance than a time-varying \gls{SRO}.
The large errors reported for DXCP-PhaT stem from recordings where in most cases the secondary \gls{GCPSD} function is calculated from frames corresponding to two different source positions.
In contrast to \textit{Scenario-3}, there are no speech pauses in \textit{Scenario-4}, when the speaker position changes. Although \gls{DWACD} can no longer use the \gls{SAD} to skip segments during position changes it shows only a small performance degradation.

The dependency of the performance of the different \gls{SRO} estimators on the standard deviation  $\sigma_\varepsilon$ of the ground truth SRO process is presented in Tab.~\ref{tab:dyn_sro}. 
For the  \gls{DWACD} method the \gls{SRO} error is nearly constant despite the growing standard deviation $\sigma_\varepsilon$. In contrast, for the online \gls{WACD} method and DXCP-PhaT
 the \gls{SRO} estimation error is increasing with $\sigma_\varepsilon$.

\begin{figure}[t]
	\centering
	\LARGE
	\resizebox{\linewidth}{!}{
		\input{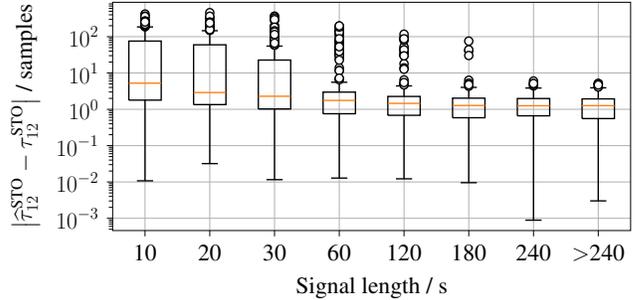}
	}
	\caption{Influence of the signal length 
		 on the \gls{STO} estimation error}
	\label{fig:sto_results}
\end{figure}
In Fig.~\ref{fig:sto_results} the distribution of the absolute \gls{STO} error is presented for different signal lengths  used for \gls{STO} estimation.
As expected, the \gls{STO} error gets smaller with growing signal lengths.
In most cases, a \SI{1}{\minute} long signal is sufficient to achieve a \gls{STO} error smaller than \SI{10}{samples}.
However there are some outliers.
These outliers disappear for signals which are at least \SI{4}{\minute} long.

\section{Conclusions}
\label{sec:conclusions}
In this paper, we present methods for \gls{SRO} and \gls{STO} estimation in scenarios with time-varying \glspl{SRO} and source position changes.
The former is caused by the time-varying deviation of the  sampling frequencies from the nominal sampling frequency, e.g., due to temperature and supply voltage changes, and is modeled as an Ornstein-Uhlenbeck process, while the latter is typical of a meeting scenario with multiple speakers engaged in a communication.
It is shown that previously proposed online \gls{SRO} estimators fail to properly handle the considered dynamic scenario, being particularly vulnerable to  source position changes.
The presented \gls{DWACD} algorithm which results from modifying our previously proposed online \gls{WACD} method handles these aspects by utilizing shorter signal segments for coherence drift estimation in combination with a more robust \gls{GCC}-based method to estimate the \gls{SRO} from a temporally weighted average coherence drift.
Furthermore, an \gls{STO} estimation method is presented which uses source-microphone distance estimates to separate the influence of the \gls{STO} from the \glspl{TDOF}. Thus, the \gls{STO} can be removed, while at the same time regarding the geometric arrangement of the WASN.  
	
	\paragraph*{Acknowledgment}
	Funded by the Deutsche Forschungsgemeinschaft (DFG, German Research Foundation) - Project 282835863.
	
	\bibliographystyle{IEEEbib}
	\bibliography{refs,library,asn_publications,p1,eulib}

\begin{thebibliography}{10}

\bibitem{Markovich-Golan2012}
Shmulik Markovich-Golan, Sharon Gannot, and Israel Cohen,
\newblock ``{Blind Sampling Rate Offset Estimation and Compensation in Wireless
  Acoustic Sensor Networks with Application to Beamforming},''
\newblock in {\em Proc. International Workshop on Acoustic Echo and Noise
  Control (IWAENC)}, 2012, pp. 1--4.

\bibitem{Miyabe2013}
Shigeki Miyabe, Nobutaka Ono, and Shoji Makino,
\newblock ``{Blind compensation of inter-channel sampling frequency mismatch
  with maximum likelihood estimation in STFT domain},''
\newblock in {\em IEEE International Conference on Acoustics, Speech and Signal
  Processing (ICASSP)}, 2013, pp. 674--678.

\bibitem{Bahari2017}
Mohamad~Hasan Bahari, Alexander Bertrand, and Marc Moonen,
\newblock ``{Blind Sampling Rate Offset Estimation for Wireless Acoustic Sensor
  Networks Through Weighted Least-Squares Coherence Drift Estimation},''
\newblock {\em IEEE/ACM Transactions on Audio, Speech, and Language
  Processing}, vol. 25, no. 3, pp. 674--686, 2017.

\bibitem{SchHaeb2017}
Joerg Schmalenstroeer, Jahn Heymann, Lukas Drude, Christoph Boeddecker, and
  Reinhold Haeb-Umbach,
\newblock ``Multi-stage coherence drift based sampling rate synchronization for
  acoustic beamforming,''
\newblock in {\em 19th International Workshop on Multimedia Signal Processing
  (MMSP)}, 2017.

\bibitem{Araki19}
Shoko Araki, Nobutaka Ono, Keisuke Kinoshita, and Marc Delcroix,
\newblock ``{Estimation of Sampling Frequency Mismatch between Distributed
  Asynchronous Microphones under Existence of Source Movements with Stationary
  Time Periods Detection},''
\newblock in {\em IEEE International Conference on Acoustics, Speech and Signal
  Processing (ICASSP)}, 2019, pp. 785--789.

\bibitem{Chinaev2021}
Aleksej Chinaev, Gerald Enzner, Tobias Gburrek, and Joerg Schmalenstroeer,
\newblock ``{Online Estimation of Sampling Rate Offsets in Wireless Acoustic
  Sensor Networks with Packet Loss},''
\newblock in {\em 29th European Signal Processing Conference (EUSIPCO)}, 2021,
  pp. 1--5.

\bibitem{Gannot2016Consolidated}
Sharon Gannot, Emmanuel Vincent, Shmulik Markovich-Golan, and Alexey Ozerov,
\newblock ``A consolidated perspective on multi-microphone speech enhancement
  and source separation,''
\newblock {\em IEEE/ACM Transactions on Audio, Speech and Language Processing},
  vol. 25, no. 4, pp. 692--730, 2016.

\bibitem{SchJebHaeb2014}
Joerg Schmalenstroeer, Patrick Jebramcik, and Reinhold Haeb-Umbach,
\newblock ``A combined hardware\hbox{-}software approach for acoustic sensor
  network synchronization,''
\newblock {\em Signal Processing}, vol. 107, pp. 171 -- 184, 2015.

\bibitem{DRIFT2021}
``High-precision audio drift measurements with gps,''
  {https://protyposis.net/clockdrift/high-precision-audio-drift-measurements-with-gps/},
\newblock Aug. 2021.

\bibitem{Walls92}
Fred~L. Walls and Jean-Jacques Gagnepain,
\newblock ``Environmental sensitivities of quartz oscillators,''
\newblock {\em IEEE transactions on ultrasonics, ferroelectrics, and frequency
  control}, vol. 39, pp. 241--9, 02 1992.

\bibitem{QSL2021}
``Sample rate and frequency calibration,''
  {https://www.qsl.net/dl4yhf/speclab/frqcalib.htm/ \\
  soundcard\_clock\_drift\_measurements},
\newblock Aug. 2021.

\bibitem{Uhlenbeck}
George~E. Uhlenbeck and Leonard~S. Ornstein,
\newblock ``On the theory of the brownian motion,''
\newblock {\em Phys. Rev.}, vol. 36, pp. 823--841, Sep 1930.

\bibitem{Gburrek21b}
Tobias Gburrek, Joerg Schmalenstroeer, and Reinhold Haeb-Umbach,
\newblock ``On source-microphone distance estimation using convolutional
  recurrent neural networks,''
\newblock in {\em Proc. 14th ITG-Symposium Speech Communication}, 2021.

\bibitem{wang16}
Lin Wang and Simon Doclo,
\newblock ``{Correlation Maximization-Based Sampling Rate Offset Estimation for
  Distributed Microphone Arrays},''
\newblock {\em IEEE/ACM Transactions on Audio, Speech, and Language
  Processing}, vol. 24, no. 3, pp. 571--582, 2016.

\bibitem{knapp76}
Charles~H. Knapp and G.~Clifford Carter,
\newblock ``{The generalized correlation method for estimation of time
  delay},''
\newblock {\em IEEE Transactions on Acoustics, Speech, and Signal Processing},
  vol. 24, no. 4, pp. 320--327, 1976.

\bibitem{fischler81}
Martin~A. Fischler and Robert~C. Bolles,
\newblock ``Random sample consensus: A paradigm for model fitting with
  applications to image analysis and automated cartography,''
\newblock {\em Commun. ACM}, vol. 24, no. 6, pp. 381–395, June 1981.

\bibitem{TCXO14}
``{TCXO} frequency stability and frequency accuracy budget,'' {SiTime,
  SiT-AN10039 Rev 1.1},
\newblock Jul. 2014.

\bibitem{gburrek2021geometry}
Tobias Gburrek, Joerg Schmalenstroeer, and Reinhold Haeb-Umbach,
\newblock ``Geometry calibration in wireless acoustic sensor networks utilizing
  doa and distance information,''
\newblock {\em EURASIP Journal on Audio, Speech, and Music Processing}, vol.
  2021, no. 1, pp. 1--17, 2021.

\bibitem{timit}
John~S. Garofolo, Lori~F. Lamel, William~M. Fisher, Jonathan~G. Fiscus,
  David~S. Pallett, Nancy~L. Dahlgren, and Victor Zue,
\newblock ``{TIMIT} acoustic-phonetic continuous speech corpus,'' 1993,
\newblock Linguistic Data Consortium (LDC).

\bibitem{Schmalen2018efficient}
Joerg Schmalenstroeer and Reinhold Haeb-Umbach,
\newblock ``{Efficient Sampling Rate Offset Compensation - An {Overlap-Save}
  Based Approach},''
\newblock in {\em 26th European Signal Processing Conference (EUSIPCO)}, 2018,
  pp. 499--503.

\bibitem{Chinaev21_DXCPPhaT}
Aleksej Chinaev, Philipp Th{\"u}ne, and Gerald Enzner,
\newblock ``{Double-Cross-Correlation Processing for Blind Sampling-Rate and
  Time-Offset Estimation},''
\newblock {\em IEEE/ACM Transactions on Audio, Speech, and Language
  Processing}, vol. 29, pp. 1881--1896, 2021.

\end{thebibliography}
	
\end{document}